# Examination of the concept of degree of rate control by first-principles kinetic Monte Carlo simulations

Hakim Meskine[a], Sebastian Matera[a], Matthias Scheffler[a,b,c],

Karsten Reuter[a], and Horia Metiu[a,b]

[a]Fritz-Haber-Institut der Max-Planck-Gesellschaft, Faradayweg 4-6, D-14195 Berlin, Germany, and [b]Department of Chemistry & Biochemistry and [c]Materials Department, University of California at Santa Barbara, CA. 93106, USA

**Abstract:** The conceptual idea of degree of rate control (DRC) approaches is to identify the "rate limiting step" in a complex reaction network by evaluating how the overall rate of product formation changes when *a small change* is made in one of the kinetic parameters. We examine two definitions of this concept by applying it to first-principles kinetic Monte Carlo simulations of the CO oxidation at $RuO_2(110)$. Instead of studying experimental data we examine simulations, because in them we know the surface structure, reaction mechanism, the rate constants, the coverage of the surface and the turn-over frequency at steady state. We can test whether the insights provided by the DRC are in agreement with the results of the simulations thus avoiding the uncertainties inherent in a comparison with experiment. We find that the information provided by using the DRC is non-trivial: It could not have been obtained from the knowledge of the reaction mechanism and of the magnitude of the rate constants alone. For the simulations the DRC provides furthermore guidance as to which aspects of the reaction mechanism should be treated accurately and which can be studied by less accurate and more efficient methods. We therefore conclude that a sensitivity analysis based on the DRC is a useful tool for understanding the propagation of errors from the electronic structure calculations to the statistical simulations in first-principles kinetic Monte Carlo simulations.

## I. Introduction

When dealing with mechanisms involving several elementary reactions, many kinetics textbooks discuss qualitatively the concept of the "rate limiting step". The (rarely questioned) idea or advantage of this popular concept is to reduce the complex network of many competing or concerting processes to just one supposedly crucial one. Various quantitative definitions of this concept have been introduced[1-4] and some controversy exists regarding which definition is most useful in applications[5,6] to catalysis. Here we examine two definitions which indicate how the overall rate of product formation changes when *a small change* is made in one of the kinetic parameters; different definitions change different quantities. Such definitions of the "degree of rate control" (DRC) are less likely to be useful to experimentalists who try to improve existing catalysts, because almost all changes that can be made in the laboratory will modify several kinetic parameters of the system by a significant amount. However, for simulations a DRC analysis that correctly identifies which parameters are most critical in controlling the kinetics provides not only a tool for analyzing the mechanism of a complex set of catalytic reactions, but gives also valuable guidance as to which aspects of the reaction mechanism should be treated most accurately.

To study the implications of various definitions of the DRC we use first-principles kinetic Monte Carlo (kMC) simulations of the CO oxidation at $RuO_2(110)$[7-14]. We regard these simulations as "computer experiments" and calculate how various criteria for determining the rate controlling step depend on the reaction conditions (temperature and reactant partial pressures). The use of a first-principles model ensures that we are studying a system that is not far from reality and for which we know exactly the structure and composition of the surface as a function of temperature and partial pressures. Among other quantities of interest this includes knowledge of the reaction mechanism, the rate constants, the rates with which individual reactions occur, the net rate of product formation, and the adsorbate distributions and concentrations. This insight enables us to determine in detail to what extent the different DRC criteria correctly illuminate the chemistry going on in the system.

## II. The model

The reaction mechanism in the first-principles kMC simulations of CO oxidation at $RuO_2(110)$ and its phenomenological counterpart have been described in detail in a previous article[14]. We therefore review here only the minimum necessary for understanding the present work. The surface has two binding sites for the reactants, the bridge (br) sites and the coordinatively unsaturated sites (cus), which are located on alternating rows of a square lattice. $O_2$ adsorbs dissociatively by placing two oxygen atoms on adjacent cus sites, or two oxygen atoms on adjacent br sites, or one atom on a br site and one on an adjacent cus site. Two oxygen atoms can recombine and desorb from a cus-cus, or a br-br or a cus-br pair of adjacent sites. CO can adsorb to a br or a cus site, and also desorb from either of them. There are four reactions between adsorbed CO and adsorbed O to form $CO_2$: $O^{cus} + CO^{cus}$, $O^{cus} + CO^{br}$, $O^{br} + CO^{cus}$, and $O^{br} + CO^{br}$. $CO_2$ desorbs instantly and its adsorption rate is zero. These reactions together with the site-to-site diffusion processes on the surface amount to a total of 26 elementary processes. The rate constants for all of them have been calculated by

using harmonic transition state theory with energies provided by density-functional theory.[11,12]

In any given kMC run we hold constant the partial pressures of CO and oxygen and the temperature. After some macroscopic period of time (often of the order of 0.1 seconds or longer) the system reaches a steady state: The surface coverage of O and CO and the amount of $CO_2$ produced per unit time become time independent. For the pressures and the temperatures used in the simulations the system does not have multiple steady states; therefore the same steady state is reached (for given temperature and partial pressures) regardless of the initial composition on the surface. Corresponding simulations have been performed and analyzed for a wide range of $(T, p_{O2}, p_{CO})$ conditions.[11,12,14] Not unexpectedly, high catalytic activity is only observed for a rather narrow range of gas-phase conditions, which coincides with O and CO both being present at the surface in appreciable amounts. For O-rich feed the surface is poisoned by oxygen, for CO-rich feed the surface is poisoned by CO, and little $CO_2$ is formed in each case. We correspondingly concentrate the present analysis of the DRC on two sets of gas-phase conditions: (1) CO pressures in the range $10^{-11}$ atm $< p_{CO} < 10^{-8}$ atm, with $p_{O_2} = 10^{-10}$ atm, and the temperature $T = 350$ K, and (2) CO pressures in the range 0.5 atm $< p_{CO} < 50$ atm, with $p_{O_2} = 1$ atm, and $T = 600$ K. In both case, the range of $p_{CO}$ was chosen so that at the lowest CO pressure the surface is covered by oxygen, at the highest pressure it is covered by CO and in the intermediate range both species are adsorbed and react efficiently to produce $CO_2$.

## III. Different definitions for the DRC

For a given process *i* we denote by $k_i^+$ the rate constant for the forward process *i* and by $k_i^-$ the rate constant of the corresponding backward process. Because of detailed balance

$$K_i = k_i^+ / k_i^- \qquad (1)$$

where $K_i$ is the equilibrium constant. In the case of adsorption-desorption we choose adsorption to be forward and desorption as backward. The oxidation reactions are irreversible and because of this the backward rate constant is zero and the equilibrium constant is infinite. It turns out that under the conditions of pressure and temperature studied here the $CO_2$ production rate is insensitive to the diffusion processes and therefore we do not further discuss them in what follows.

The efficiency of the oxidation process is described by the turn-over frequency (TOF), which is the number of $CO_2$ molecules produced per unit time, per unit area. Let us consider first the sensitivity of the TOF to changes of the activation energy of process *i*. If we change it by varying the barrier (either height or shape) as illustrated in Fig. 1a we affect both the forward rate constant $k_i^+$ and the backward rate constant $k_i^-$, but keep the equilibrium constant $K_i$ unchanged (i.e. the regions around the minima on the potential energy surface corresponding to reactants and products are not affected when we change the saddle point energy). As originally proposed by Campbell[4,6] the corresponding DRC criterion for process *i* is

$$x_i = \frac{k_i^+}{\text{TOF}} \frac{\partial \text{TOF}}{\partial k_i^+}\bigg|_{k_{j \neq i}^+, K_i}, \qquad (2)$$

where the factor in front of the partial derivative is introduced to make $x_i$ dimensionless. The subscripts $k_{j\neq i}^+$ and $K_i$ indicate that the partial derivative with $k_i^+$ is taken by keeping fixed all forward rate constants other than $k_i^+$ and by keeping fixed all equilibrium constants. Because of the detailed balance equation, Eq. (1), this implies that $k_i^+$ and $k_i^-$ vary so that $K_i$ does not change and that all backward rate constants other than $k_i^-$ are fixed.

A second way of changing the activation energy of process $i$ is illustrated in Fig. 1b and can e.g. be accomplished by varying the reactant minimum (either depth or shape). In analogy one can then define a corresponding DRC criterion of process $i$ as

$$x_i^+ = \frac{k_i^+}{\text{TOF}} \frac{\partial \text{TOF}}{\partial k_i^+}\bigg|_{k_{j\neq i}^+, K_{j\neq i}, k_i^-} . \qquad (3)$$

Here the partial derivative is thus taken by varying $k_i^+$ and keeping all other (forward and backward) rate constants fixed. This then automatically implies that the equilibrium constants $K_{j\neq i}$ of all processes other than $i$ are also fixed, whereas in contrast to the definition of $x_i$ in Eq. (2) now $K_i$ is changed. A quantity $x_i^-$ is defined similarly (interchange + and – in Eq. (3)) and it gives the sensitivity to changes in the product minimum, i.e. the backward $i$ reaction. Even though obvious generalizations to the $x_i$ suggested by Campbell, we are not aware of a previous use of these DRC criteria.

Through Eqs. (2) and (3) the different DRC criteria are well defined and can thus be measured in a simulation. By construction an analysis of these criteria will then reveal the sensitivity of the simulated TOF to corresponding changes in the different kinetic parameters. However, care has to be taken if one aspires to interpret these sensitivities furthermore in terms of the underlying potential energy surface. As already done in the motivation of the three DRC criteria above one obvious interpretation of measured sensitivities $x_i$, $x_i^+$ and $x_i^-$ of an adsorption process of one species to a specific surface site type would e.g. be that they reflect the sensitivities to a change in the activation barrier (and hence the local sticking coefficient at the surface site type[12]), to a change in the partial pressure of the species and to a change in the binding energy of the species to the surface site type, respectively. However, if there are several site types at the surface as in the presently studied system, it is physically not meaningful to analyze the sensitivity of the system to just a change of the species partial pressure and thus impingement to one site type; such a situation cannot be realized. Equally, a change in the binding energy of a given reaction participant at one specific site will in general affect several rate constants simultaneously, whereas the measured $x_i^-$ just reflect the sensitivity to changes of individual rate constants. For example, $CO^{cus}$ is involved in the following reactions:

$$CO^{gas} \underset{k_1^-}{\overset{k_1^+}{\rightleftarrows}} CO^{cus} \qquad (4)$$

$$CO^{cus} + O^{cus} \xrightarrow{k_2^+} CO_2 \qquad (5)$$

$$CO^{cus} + O^{br} \xrightarrow{k_3^+} CO_2 \qquad (6)$$

A change in the binding energy of $CO^{cus}$ will thus change the activation energy in $k_1^-$, $k_2^+$ and $k_3^+$, as well as that of the diffusion rate constants out of a cus site onto the neighboring cus and bridge sites. Either $x_1^-$, $x_2^+$ or $x_3^+$ measure, however, only the change of the TOF when one of these kinetic parameters is changed and can thus not directly be related to a change in the binding energy of $CO^{cus}$. If such interdependencies are kept in mind, it may still be possible to establish such a relationship by viewing the individual DRC criteria in a linear response sense as "building blocks" to the real sensitivity of the system upon a simultaneous change in several rate constants. In this respect, one would arrive at an assessment of the sensitivity to the binding energy of CO$^{cus}$ from a joint inspection of the three measured $x_1^-$, $x_2^+$ and $x_3^+$ (as well as of the DRCs of the diffusion processes).

In general, it is in the same sense and with the same caveats useful to attempt an interpretation of measured sensitivities $x_i$, $x_i^+$ and $x_i^-$ of a reaction process of two reactants to form an adsorbed product as reflecting the sensitivities to a change in the activation barrier, to a change in the binding energies of the adsorbed reactants and to a change in the binding energy of the adsorbed product, respectively. However, for the irreversible CO oxidation reactions in the specific model studied here an examination of $x_i$ and $x_i^-$ is meaningless, and the straightforward interpretation of the measured sensitivities $x_i^+$ of the CO oxidation processes is instead in terms of a change in the activation barrier (through either of the modifications shown in Fig. 1a or Fig. 1b).

For very simple reaction schemes the different DRC criteria can be evaluated analytically. However, in general they must be determined numerically. To do this we vary the rate constant of interest in very small increments around its correct value, calculate the changes in the TOF, fit the results to a polynomial, and take the linear term to be the desired derivative. With the definitions given above one has

$$x_i = x_i^+ + x_i^- \qquad . \qquad (7)$$

If all the terms are calculated independently this relationship can thus be exploited to test the numerical computations. Alternatively, one may use this relationship to calculate $x_i$ from $x_i^+$ and $x_i^-$, which is what we do consistently below. In a similar sense it is possible to exploit $\sum_i x_i = 1$ as another useful relation for the numerical calculations.

## IV. A DRC analysis of steady-state catalysis

Figure 2 shows the dependence of the steady-state TOF and the surface coverages on the partial pressure of CO, when the temperature is 350 K and the partial pressure of $O_2$ is $10^{-10}$ atm. As apparent from the figure the CO partial pressure range shown comprises the oxygen poisoned situation at the lowest $p_{CO}$, passes through the state of most efficient CO oxidation catalysis with a coexistence of both reactants at the surface at intermediate $p_{CO}$, and ends with the CO poisoned situation at the highest $p_{CO}$.[11,12,14] When we turn to the computed DRC criteria in this pressure range, the first remarkable observation is that out of the five adsorption processes, four reaction processes and six forward diffusion processes in the model only the three processes shown in the upper panel of Fig. 3 have an appreciable $x_i$ somewhere in this wide range of gas-phase conditions. There are thus quite a number of kinetic parameters that never play an important role in the overall reaction network for any of the three distinctly different and representative states of the system, i.e. O-poisoned, CO-poisoned and catalytically most active coexistence regime. In the prior two regimes there is in fact each time really only one step left that predominantly controls the catalytic activity, namely the adsorption of CO onto a cus site and the adsorption of oxygen onto a neighboring pair of cus sites, respectively. On the other hand this is a different "rate limiting step" in the two regimes and particularly in the most relevant catalytically most active state of the system it is not one, but the group of three processes that determines the overall TOF. A simplification of the reaction network by exploiting the bottleneck function of just one rate limiting process as frequently discussed in the literature would therefore not be permissible, if one aspires a correct description of the entire range of environmental conditions shown in Fig. 3.

Remarkably, one arrives at essentially the same conclusions when analyzing the alternative DRC criteria $x_i^+$ and $x_i^-$ shown in the bottom panel of Fig. 3, as well as when analyzing the completely different set of gas-phase conditions shown in Fig. 4. For $T = 600$K and a fixed partial pressure of $O_2$ of 1 atm, the partial pressure of CO is there varied in the range 0.5 atm $< p_{CO} <$ 50 atm, thereby covering again the O-poisoned regime at the lowest $p_{CO}$, the state of most efficient CO oxidation catalysis with a coexistence of both reactants at the surface at intermediate $p_{CO}$, and the CO poisoned situation at the highest $p_{CO}$.[11,12,14]

We begin a closer examination of the chemistry revealed by the different DRC criteria by focusing on the computed $x_i$ for the oxygen poisoned regime on the left hand side in Fig. 3. Under these conditions this DRC criterion tells that predominantly the adsorption and desorption process of CO into and out of cus sites ($CO^{gas} \leftrightarrow CO^{cus}$) and to a smaller extent the adsorption and desorption of $O_2$ into and out of a pair of cus sites ($O_2^{gas} \leftrightarrow O^{cus}/O^{cus}$) is controlling the catalytic activity. More specifically, an increase in the $CO^{cus}$ adsorption rate constant under a fixed $CO^{gas} \leftrightarrow CO^{cus}$ equilibrium constant would increase the TOF, while a decrease in the $O_2^{cus/cus}$ adsorption rate constant under a fixed $O_2^{gas} \leftrightarrow O^{cus}/O^{cus}$ equilibrium constant would slightly decrease the TOF. Small variations in the four O+CO reaction rate constants, on the other hand, would for example not much influence the TOF in this regime; neither would this be the case for any process involving the bridge sites. This focus on the $CO^{gas} \leftrightarrow CO^{cus}$ and $O_2^{gas} \leftrightarrow O^{cus}/O^{cus}$ processes is also carved out by the DRC criteria $x_i^+$ and $x_i^-$ in the bottom panel of Fig. 3, which now, however, distinguish between changes due to a variation of the forward (adsorption) and backward (desorption) rate constant without conserving the equilibrium constant. Interestingly, for corresponding changes the adsorption and desorption of cus oxygen turns out to be equally important as the adsorption of CO$^{\text{cus}}$, whereas changes as monitored by the $x_i$ DRC clearly put the emphasis on the CO$^{\text{cus}}$ process alone.

The analysis of these findings based on the detailed data provided by our first-principles kMC "computer experiments"[11,12,14] reveals that the complementary information provided by the two types of DRC criteria nicely carves out the chemistry of the system under these gas-phase conditions: In this O-poisoned regime any change that leads to an increased presence of CO at the surface will be favorable for the catalytic activity, just as much as will be any change that leads to a decreased presence of surface oxygen. More specifically, this holds primarily for the cus sites, since the very strong binding energy of oxygen at the bridge sites leads to an almost complete deactivation of these sites under these gas-phase conditions. This understanding is fully consistent with the processes that are identified by the two types of DRC criteria.

The presence of a species at the surface is determined by the balance between accumulation due to adsorption on the one hand and depletion due to desorption and reaction on the other. For the predominant O$^{\text{cus}}$ species the small depopulation resulting from the few reaction events that take place at the lowest $p_{\text{CO}}$ in the pressure range on the left hand side of Fig. 3 hardly plays any role. Correspondingly its presence at the surface is primarily governed by the competition between adsorption and desorption. Changing the adsorption rate constant has then basically the same inverse effect as varying the desorption rate constant, as correctly picked up by the $x_i^+$ and $x_i^-$ for the O$_2^{\text{gas}}$ ↔ O$^{\text{cus}}$/O$^{\text{cus}}$ process, which exhibit similar magnitude and opposing sign at the lowest $p_{\text{CO}}$. In this situation, changes of this process that conserve the equilibrium constant and thus equally favor or disfavor both the forward and backward reaction do not much affect the presence at the surface, and correspondingly the $x_i$ computed for the $O_2^{gas} \leftrightarrow O^{cus}/O^{cus}$ process is close to zero. With increasing CO partial pressure and thereby increasing TOF, the role played by the increasing number of reaction processes for the depopulation of $O^{cus}$ becomes more and more important and the symmetry in the relevance of adsorption and desorption for the presence of $O^{cus}$ at the surface is lost. Again, this is nicely reflected by the $x_i^+$ and $x_i^-$ of the $O_2^{gas} \leftrightarrow O^{cus}/O^{cus}$ process, which still have opposite sign but start to deviate from another in magnitude. In this skewed situation, now also modifications of the $O_2^{gas} \leftrightarrow O^{cus}/O^{cus}$ process that equally affect adsorption and desorption begin to carry through to the $O^{cus}$ surface presence and the corresponding $x_i$ DRC criterion starts to exhibit non-zero values.

A different scenario holds for the minority $CO^{cus}$ species at these low $p_{CO}$ gas-phase conditions. For this species the few reaction events are essentially the main depopulation channel, since they have a higher rate constant than the desorption process, i.e. $CO^{cus}$ at the surface are rather removed by the $O^{cus}+CO^{cus}$ reaction than by desorption. Slight changes in the desorption rate constant are then not significant for the presence of $CO^{cus}$ and therewith for the total TOF, as properly reflected by the small $x_i^-$ of the $CO^{gas} \leftarrow CO^{cus}$ process. There is thus no symmetry between adsorption and desorption in determining the presence of $CO^{cus}$ at the surface, so that the latter can also be changed by modifications of the $CO^{gas} \leftrightarrow CO^{cus}$ process that conserve the equilibrium constant. The latter are thereby also of relevance for the TOF as correctly indicated by the large corresponding $x_i$ DRC criterion for the $CO^{gas} \leftrightarrow CO^{cus}$ process in the upper panel of Fig. 3.

Having established that the insights provided by the DRC criteria are in full agreement with the detailed information we have from the simulations for the O-poisoned regime on the left hand side of Fig. 3, we proceed by examining how we can also use these criteria to obtain guidance as to which aspects of the reaction mechanism are most important to obtain a quantitatively correct description under these gas-phase conditions. The large $x_i$ for the $CO^{gas}$ ↔ $CO^{cus}$ process tells that changes of the adsorption rate constant that conserve the equilibrium constant have a large impact on the catalytic activity. One immediate microscopic quantity that leads to such a change is the local sticking coefficient[12] for adsorption of CO into the cus sites, and we thus learn that uncertainties in this computed quantity will directly propagate to the mesoscopic kMC simulation results. Other local sticking coefficients describing the adsorption of CO at bridge sites or oxygen at cus or bridge site pairs, on the other hand, are not that critical for the proper description of the system in this O-poisoned state.

A similar analysis of the large $x_i^-$ for the $O^{cus}/O^{cus}$ desorption process, reflecting a change where the backward desorption rate constant is changed without conserving the equilibrium constant, would suggest the binding energy of oxygen at the cus sites as another crucial microscopic quantity. Such an interpretation is, however, a typical example that reveals the limitations of the employed DRC criteria that are only sensitive to changes that are made with respect to a single rate constant. Changing the $O^{cus}$ binding energy would not only affect the desorption out of this site (both $O^{cus}/O^{cus}$ and $O^{cus}/O^{br}$ pairs), but also all reaction channels involving this species ($O^{cus}$ + $CO^{cus}$, $O^{cus}$+$CO^{br}$), as well as diffusion processes of this species. While it would therefore not be permissible to conclude in general from the large computed $x_i^-$ for the $O^{cus}/O^{cus}$ desorption process on the importance of the $O^{cus}$ binding energy, this seems justified under these specific gas-phase conditions, since none of the involved other processes exhibits a non-zero DRC criterion, cf. Fig. 3. Likewise, it is also possible to conclude from the small values of the computed $x_i^-$ DRC criteria on a low sensitivity of the simulated TOF in this pressure range on the binding energies of oxygen at bridge sites, as well as of CO at bridge and cus sites.

Finally, the straightforward interpretation for the large $x_i^+$ found in this gas-phase regime for the adsorption of oxygen and CO at cus sites would be the effect of pressure on the TOF. A change in partial pressure affects the forward (adsorption) rate constant without conserving the equilibrium constant and is thus exactly a change that is picked up by this DRC criterion. Again, in a multi-site system such as the one studied here one has to be careful with this interpretation, since it is obviously physically not possible to change partial pressures in such a way that only the impingement on one site type is affected. For the present conditions, however, this is not much of a problem due to the inactive role of the bridge sites (as reflected by the zero $x_i^+$ of adsorption processes into bridge sites). This allows to make the meaningful, but maybe not too enlightening interpretation that the large positive $x_i^+$ for adsorption of CO at cus sites reveals that increasing the CO partial pressure will increase the catalytic activity, since it helps to bring the system further out of the O-poisoned towards the catalytically active coexistence state. Similarly, the large negative $x_i^+$ for adsorption of oxygen at cus site pairs reveals that increasing the oxygen partial pressure is detrimental for the catalysis, since it drives the system even further into the O-poisoned state.

Summarizing, the detailed sensitivity analysis enabled by the DRC criteria points therefore to the sticking coefficient of CO at cus sites and the $O^{cus}$ binding energy as the two quantities that are predominantly responsible for an accurate description of the catalytic activity in the O-poisoned state at the lower $p_{CO}$ in Fig. 3. A completely equivalent line of analysis as the one just detailed shows that the DRC criteria also fully pick up the chemistry of the system in the CO-poisoned state, i.e. at the higher end of the $p_{CO}$ range shown in Fig. 3. Now it is the presence of $O^{cus}$ at the surface that rules the catalysis, and this presence is to the largest extent determined by competing adsorption and desorption processes. Correspondingly the only largely non-zero $x_i$ belongs to the $O_2^{gas} \leftrightarrow O^{cus}/O^{cus}$ process, with a positive $x_i^+$ connected to the adsorption of oxygen into a cus site pair, a negative $x_i^+$ connected to the adsorption of CO into a cus site, and a positive $x_i^-$ connected to the desorption of CO out of a cus site. The important microscopic quantities for a proper description of this CO-poisoned state that are therefore filtered out are the sticking coefficient of oxygen at a cus site pair and the $CO^{cus}$ binding energy. The almost quantitative agreement[11,12] that was reached by the present first-principles kMC model with experimental data that largely corresponds to these gas-phase conditions suggests therefore that especially these two microscopic quantities are rather well described in the model.

Turning to the catalytically most relevant coexistence region in the middle of the pressure range shown in Fig. 3, we observe rapid variations of several DRC criteria. Particularly in the range $2 \cdot 10^{-11}$ atm $< p_{CO} < 4 \cdot 10^{-11}$ atm the adsorption-desorption related DRC seem to diverge. This is not a real divergence: At these partial pressures the variation of the TOF with the CO pressure becomes almost vertical, cf. Fig. 2, and it is difficult to numerically determine the slope of this steep increase over more than seven orders of magnitude. The large values exhibited by the $x_i$, $x_i^+$ and $x_i^-$ related to the $CO^{gas} \leftrightarrow CO^{cus}$ and $O_2^{gas} \leftrightarrow O^{cus}/O^{cus}$ processes merely reflect that already minute changes in the description of the adsorption and desorption of these species have a large effect on the amount and spatial distribution of both reactants coexisting at the surface in this regime, and therewith on the catalytic activity. It is also only in this high TOF regime that the depletion of surface species by the frequent reaction events can become comparable to the depletion due to the on-going desorption events of both species, and correspondingly it is only in the corresponding narrow range of CO partial pressures that we obtain a non-zero DRC criterion for reaction processes. In line with its predominant role for the total TOF under these gas-phase conditions it is specifically the $O^{cus}+CO^{cus}$ reaction that exhibits a large DRC in Fig. 3.

While the situation in this coexistence region is thus more complex, we still find that also here the insight provided by the different DRC criteria is in complete agreement with the knowledge we have about the system from the detailed analysis of the data available from the first-principles kMC simulations, i.e. the total TOF and the contribution to it from the various reaction mechanisms, the surface coverages, as well as the occurrence of the individual elementary processes. This agreement is quite remarkable considering that already the pressure range analyzed in Fig. 3 comprises three quite distinct and representative systems states. Even more remarkable is that the situation is exactly the same when one conducts an equivalent examination of the DRC data compiled in Fig. 4 for a completely different range of pressures at a more elevated temperature. Also here, the DRCs correctly describe the chemistry of the system. Since the story is essentially the same as the one for the $T = 350$ K data in Fig. 3, we do not elaborate on it in detail, but only note that the main difference is that at the higher temperature the desorption of $CO^{cus}$ and of $CO^{br}$ has become much faster. In case of $CO^{cus}$ it now actually competes with the $O^{cus}+CO^{cus}$ reaction as the main $CO^{cus}$ depopulation channel in the O-poisoned regime. Correspondingly, adsorption and desorption determine then more symmetrically the presence of $CO^{cus}$ at the surface. In contrast to the situation at $T = 350$ K, this leads therewith to a small $x_i$ for the $CO^{gas} \leftrightarrow CO^{cus}$ process in this regime in Fig. 4, while the increased importance of the competition between the $O^{cus}+CO^{cus}$ reaction and $CO^{cus}$ desorption for the total $CO^{cus}$ presence at the surface and thus the TOF is nicely reflected by the larger DRCs of this reaction and of this desorption process.

## V. The apparent activation energy

In the previous section we have shown that the DRC criteria provide useful and non-trivial insight into the chemistry of the system and can furthermore be employed to obtain guidance as to which microscopic parameters critically determine the overall catalytic activity under different gas-phase conditions. Here we show that $x_i^\sigma$ is also useful in explaining the effective activation energy of the network.

When analyzing experimental results it is common to plot the steady-state activity data in form of an Arrhenius plot, i.e. as the logarithm of TOF versus $1/T$. In many cases this yields a straight line whose slope

$$E_{app} = -\frac{\partial \ln(\text{TOF})}{\partial \beta} \qquad (8)$$

is then viewed as an "apparent" or "effective" activation energy, which is believe to convey information on the rate-limiting step in the reaction network.[15] Here $\beta = 1/k_B T$ and $k_B$ is the Boltzmann constant. While the limitations and danger of this concept are well documented in the literature,[16] it still prevails in practical research and the mere existence of a straight line in some data range is sometimes used to argue that the corresponding $E_{app}$ provides insight into a bottleneck elementary process at the corresponding gas-phase conditions.

Let us assume that an Arrhenius plot, in a certain temperature range, is a straight line so that an apparent activation barrier can be defined. Since the corresponding steady-state TOF is a function of the rate constants of all elementary processes in the system we have

$$E_{app} = -\sum_{i,\sigma}\left(\frac{\partial \ln(\text{TOF})}{\partial \ln k_i^\sigma}\right)_{k_j^\sigma} \frac{\partial \ln k_i^\sigma}{\partial \ln \beta} \quad , \tag{9}$$

where the sum runs over all forward and backward elementary processes with rate constants $k_j^\sigma$ (and $\sigma = +$ or $-$), and the subscript $k_j^\sigma$ indicates that when taking the derivative with one rate constant one keeps all other rate constants fixed. Note that there is no explicit dependence of TOF on the steady-state surface coverages, since the latter are themselves functions of the underlying rate constants, and we have furthermore assumed that under steady-state conditions there is no dependence on the initial state of the system either since there are no multiple steady states. If we further assume that the rate constants can be written in an Arrhenius type form

$$k_i^\sigma = f_i^\sigma(\beta, p_i) \exp[-\Delta E_i^\sigma \beta] \quad , \tag{10}$$

where the pre-exponentials $f_i^\sigma(\beta, p_i)$ are weakly dependent on $\beta$ and on the partial pressures $p_i$, we can use the definition of $x_i^\sigma$ given in Eq. (3) to rewrite Eq. (9) as

$$E_{app} = \sum_{i,\sigma} x_i^\sigma \Delta E_i^\sigma - \sum_{i,\sigma}\left(\frac{\partial \ln f_i^\sigma}{\partial \ln \beta}\right)\left(\frac{\partial \ln(\text{TOF})}{\partial \ln k_i^\sigma}\right)_{k_j^\sigma} \quad . \tag{11}$$

Since the pre-exponential $f_i^\sigma$ is weakly dependent on $\beta$, the derivative in the second term is small and to a good approximation we have

$$E_{app} \approx \sum_{i,\sigma} x_i^\sigma \Delta E_i^\sigma \quad . \tag{12}$$

We therefore arrive at the result that, under the set of assumptions made, the apparent activation energy is approximately given by an "average" of the activation energies of all elementary processes weighted with their DRC criteria $x_i^\sigma$. Recalling that $\sum_i x_i = 1$, this shows that if there is only one process with an appreciable DRC, then $E_{app}$ indeed roughly reveals the activation energy of this bottleneck process. However, as illustrated in the preceding section, there is no reason to expect that such a situation is general. In the case of multiple rate controlling processes Eq. (12) offers then an interpretation of the effective activation energy.

Since there are no multiple steady states in the present kMC model and the underlying first-principles rate constants derived via transition state theory can be expressed in an Arrhenius like form, we can illustrate this using simulated TOFs again as a "computer experiment". In the upper panel of Fig. 5 we plot ln(TOF) kMC "data" versus $1/T$, for fixed partial pressures of $p_{O_2}$ = 1 atm and $p_{CO}$ = 2 atm. We obtain a straight line in the temperature range between 350 K and 500 K, which is in the regime where the surface is almost entirely covered with CO. From the graph in that temperature range we obtain an apparent activation energy of $E_{app}$ = 2.85 eV, by fitting a straight line to the kMC data.

We know the activation energies of all elementary processes in our kMC simulation and none is even close to 2.85 eV, which demonstrates immediately that the deduced apparent activation energy does not reflect the activation energy of one bottleneck process. Indeed, in the temperature range in which the straight line fitting was performed there are instead three processes having a sizeable $x_i^\sigma$: The adsorption of CO on the cus sites, the desorption of CO from the cus sites and the dissociative adsorption of oxygen into a pair of cus sites. Of these, only the desorption of CO$^{cus}$ is activated, having a barrier of 1.3 eV[11,12]. Since $x_i^-$ for this process is ~2 throughout the temperature range of interest, Eq. (12) gives $E_{app}$ ~ 2.6 eV, which considering the approximations made is fairly close to the true value of 2.85 eV.

## VI. Summary

We have used first-principles kinetic Monte Carlo simulations to study the usefulness of two definitions of the DRC: One given by Campbell[4,6] and one defined here. Both definitions study the "linear response" of the turn-over frequency to a change in one of the rate constants of the reaction network. Analyzing the complementary insight provided by the two definitions over a wide range of gas-phase conditions we conclude that they correctly reflect the knowledge we have about the system from the detailed data available from the first-principles kMC simulations, i.e. the total TOF and the contribution to it from the various reaction mechanisms, the surface coverages, as well as the occurrence of the individual elementary processes. The conclusions reached by calculating the DRC are furthermore non-trivial in the sense that they could not have been reached by merely examining the magnitude of the rate constants or of the activation energies of the elementary processes.

In the pressure regime in which the catalyst is most active the DRC analysis identifies an entire group of processes to which the TOF is very sensitive. While there is thus no single "rate limiting step" this number of processes controlling the overall CO$_2$ production is small. This indicates that if the rate constants of these processes are known accurately, the kMC procedure will produce correct results even if the other rate constants are inaccurate. We have confirmed this by direct calculation of the variation of the TOF with some of the unimportant rate constants. In some cases one can change a rate constant by several orders of magnitude with no effect on the TOF. Apart from providing a tool for analyzing the mechanism of a complex set of catalytic reactions, the DRC tells us therefore which aspects of the reaction mechanism must be treated accurately and which can be studied by less accurate and more efficient methods. In this sense we argue that a sensitivity analysis based on the DRC can be a useful tool towards establishing a control of the propagation of error from the electronic structure calculations to the statistical simulations in first-principles kinetic Monte Carlo approaches.

**Figure captions:**
**Fig. 1.** Schematic illustration of small changes of the potential energy surface that would correspond to the two definitions of the DRC (see text).

**Fig. 2.** Dependence of the steady-state TOF (upper panel) and site occupations (lower panel) obtained with the first-principles kMC simulations for $T = 350$ K and $p_{O_2} = 10^{-10}$ atm. Shown in the upper panel is the dependence of the total TOF, as well as the contribution of the four different reaction mechanisms. The lower panel shows the average occupation of the bridge and cus sites by O or CO.

**Fig. 3.** Dependence of $x_i$ (upper panel) and $x_i^\sigma$ (lower panel) on the CO partial pressure during steady-state for the same set of gas-phase conditions as in Fig. 2, i.e. $T = 350$ K and $p_{O_2} = 10^{-10}$ atm. See text for an explanation of the nomenclature used to describe the different elementary processes. The DRC values for all processes not shown are practically zero on the scale of this figure. Due to the irreversibility of the CO oxidation reactions in the model we only show the DRC $x_i^+$ in the upper panel for clarity.

**Fig. 4.** Same as Fig. 3, but now for gas-phase conditions with $T = 600$ K and $p_{O_2} = 1$ atm.

**Fig. 5.** Upper panel: Plot of the logarithm of the simulated steady-state turnover frequency for $CO_2$ production versus the inverse temperature $(1/k_B T)$ for $p_{O_2} = 1$ atm and $p_{CO} = 2$ atm. Lower panel: Computed $x_i$ and $x_i^\sigma$ for these gas-phase conditions. See text for an explanation of the nomenclature used to describe the different elementary processes. The DRC values for all processes not shown are practically zero on the scale of this figure.

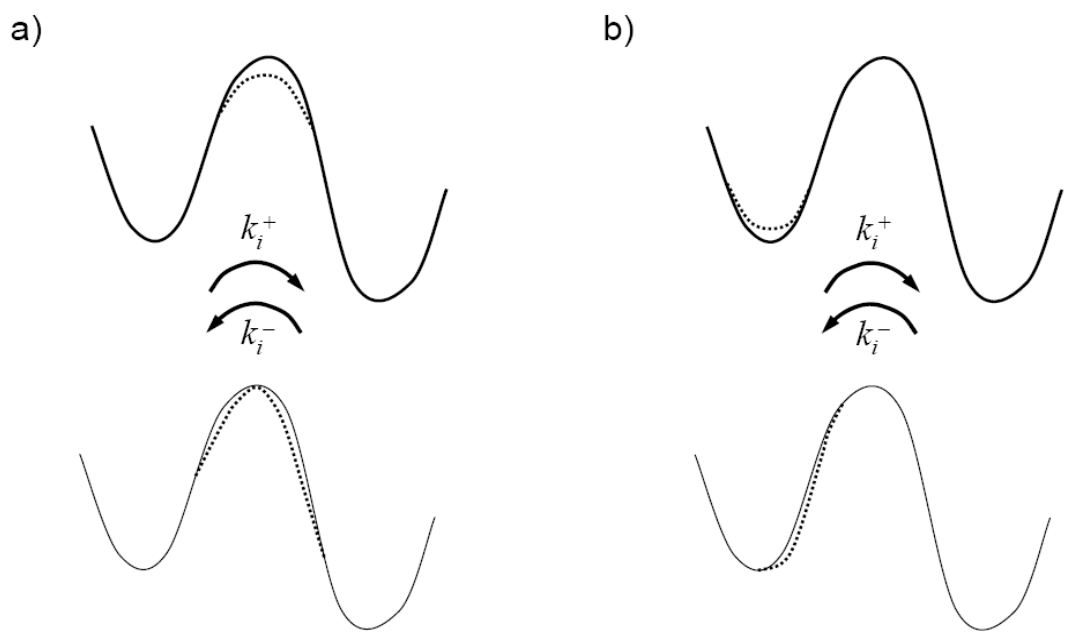

**Fig. 1.** Schematic illustration of small changes of the potential energy surface that would correspond to the two definitions of the DRC (see text).

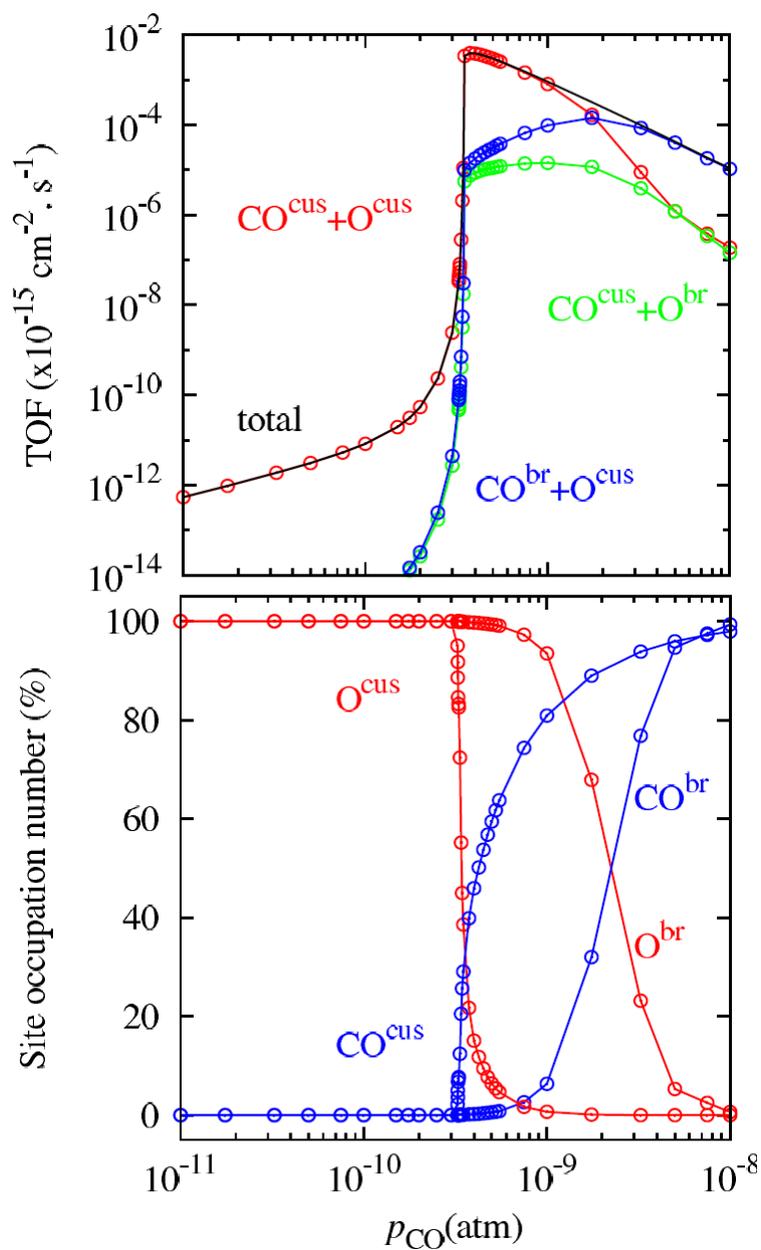

**Fig. 2.** Dependence of the steady-state TOF (upper panel) and site occupations (lower panel) obtained with the first-principles kMC simulations for $T = 350$ K and $p_{O_2} = 10^{-10}$ atm. Shown in the upper panel is the dependence of the total TOF, as well as the contribution of the four different reaction mechanisms. The lower panel shows the average occupation of the bridge and cus sites by O or CO.

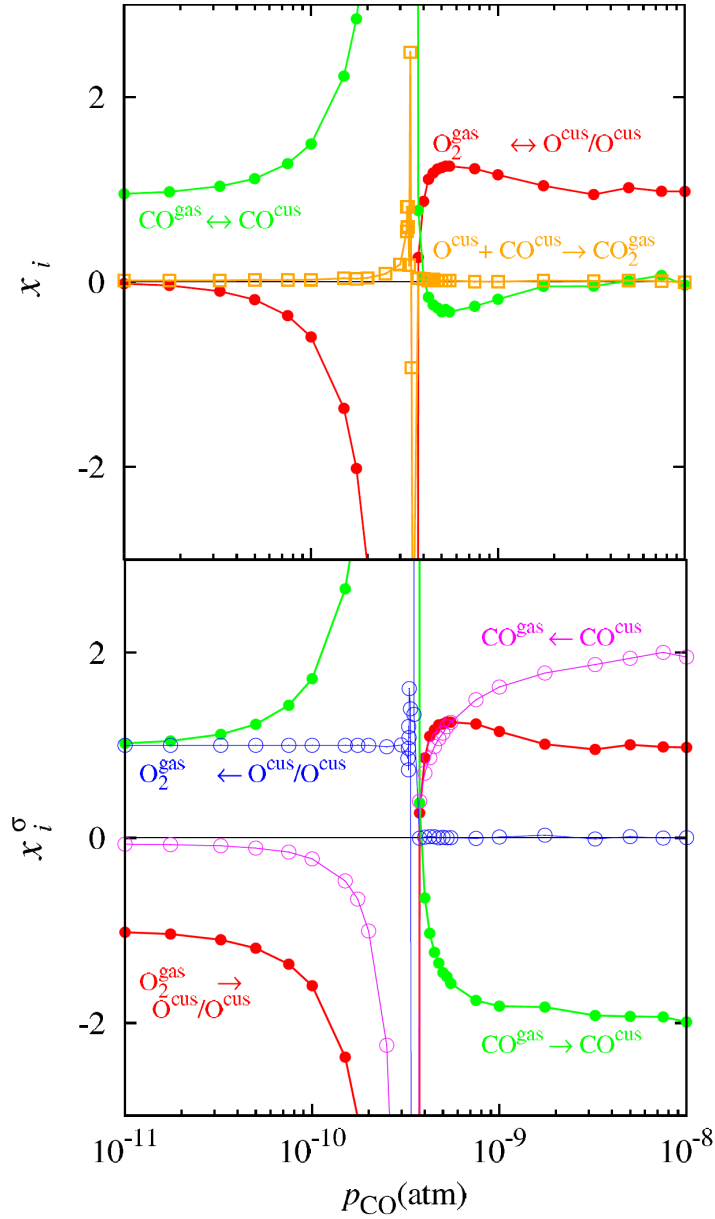

**Fig. 3.** Dependence of $x_i$ (upper panel) and $x_i^\sigma$ (lower panel) on the CO partial pressure during steady-state for the same set of gas-phase conditions as in Fig. 2, i.e. $T = 350$ K and $p_{O_2} = 10^{-10}$ atm. See text for an explanation of the nomenclature used to describe the different elementary processes. The DRC values for all processes not shown are practically zero on the scale of this figure. Due to the irreversibility of the CO oxidation reactions in the model we only show the DRC $x_i^+$ in the upper panel for clarity.

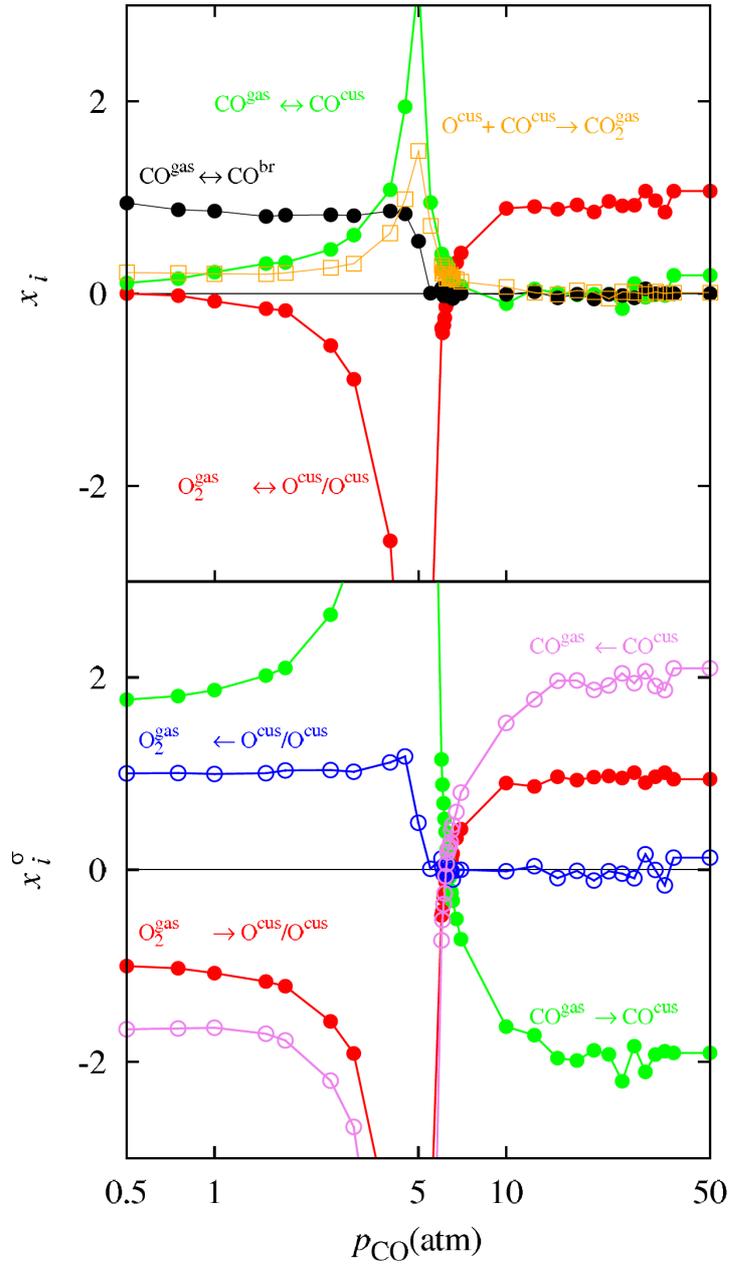

**Fig. 4.** Same as Fig. 3, but now for gas-phase conditions with $T = 600$ K and $p_{O_2} = 1$ atm.

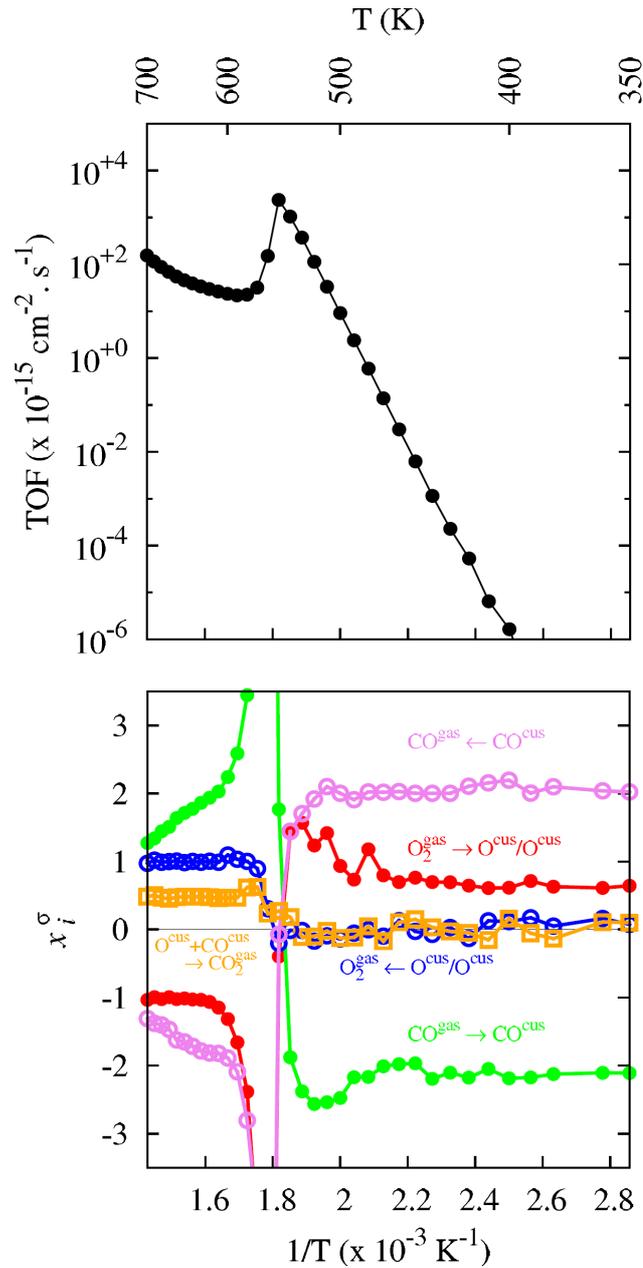

**Fig. 5.** Upper panel: Plot of the logarithm of the simulated steady-state turnover frequency for $CO_2$ production versus the inverse temperature ($1/k_B T$) for $p_{O_2} = 1$ atm and $p_{CO} = 2$ atm. Lower panel: Computed $x_i$ and $x_i^\sigma$ for these gas-phase conditions. See text for an explanation of the nomenclature used to describe the different elementary processes. The DRC values for all processes not shown are practically zero on the scale of this figure.